\documentclass[english,aps,prl,superscriptaddress,twocolumn]{revtex4}
\usepackage[T1]{fontenc}
\usepackage[latin9]{inputenc}
\usepackage{textcomp}
\usepackage{graphicx}

\makeatletter
\@ifundefined{textcolor}{}
{%
 \definecolor{BLACK}{gray}{0}
 \definecolor{WHITE}{gray}{1}
 \definecolor{RED}{rgb}{1,0,0}
 \definecolor{GREEN}{rgb}{0,1,0}
 \definecolor{BLUE}{rgb}{0,0,1}
 \definecolor{CYAN}{cmyk}{1,0,0,0}
 \definecolor{MAGENTA}{cmyk}{0,1,0,0}
 \definecolor{YELLOW}{cmyk}{0,0,1,0}
 }

\makeatother

\usepackage{babel}
\begin{document}

\title{Reentrant valence transition in EuO at high pressures: beyond the
bond-valence model}

\author{N. M. Souza-Neto}

\email{narcizo.souza@lnls.br}

\selectlanguage{english}%

\affiliation{Advanced Photon Source, Argonne National Laboratory, Argonne, IL
60439, U.S.A.}

\affiliation{Laboratório Nacional de Luz Síncrotron, Campinas, SP 13083-970, Brazil}

\author{J. Zhao }

\affiliation{Advanced Photon Source, Argonne National Laboratory, Argonne, IL
60439, U.S.A.}

\author{E. E. Alp}

\affiliation{Advanced Photon Source, Argonne National Laboratory, Argonne, IL
60439, U.S.A.}

\author{G. Shen}

\affiliation{HPCAT, Geophysical Laboratory, Carnegie Institution of Washington,
U.S.A.}

\author{S. V. Sinogeikin}

\affiliation{HPCAT, Geophysical Laboratory, Carnegie Institution of Washington,
U.S.A.}

\author{G. Lapertot}

\affiliation{SPSMS, UMR-E 9001, CEA-INAC/UJF-Grenoble 1, 17 rue des martyrs, 38054
Grenoble, France}

\author{D. Haskel}

\email{haskel@aps.anl.gov}

\selectlanguage{english}%

\affiliation{Advanced Photon Source, Argonne National Laboratory, Argonne, IL
60439, U.S.A.}
\begin{abstract}
The pressure-dependent relation between Eu valence and lattice structure in model compound EuO is studied with synchrotron-based x-ray spectroscopic and diffraction techniques. Contrary to expectation, a 7\% volume collapse at $\approx$ 45 GPa is accompanied by a reentrant Eu valence transition into a $\emph{lower}$ valence state. In addition to highlighting the need for probing both structure and electronic states directly when valence information is sought in mixed-valent systems, the results also show that widely used bond-valence methods
fail to quantitatively describe the complex electronic valence behavior of EuO under pressure.

\end{abstract}
\maketitle
The phenomenon of mixed-valency in \emph{f}-electron systems occurs when
otherwise localized \emph{f}-orbitals of Rare-Earth or Actinide elements
hybridize in the solid with \emph{s},\emph{p},\emph{d} electrons.
A quantum superposition of differently occupied (valence) \emph{f}-states
emerges when the energy difference between competing single-valent
states is smaller than the \emph{f}-electron bandwidth, usually termed
fluctuating valence state \cite{Varma-RMP76}. The onset of mixed-valency
under applied pressure, chemical substitutions, or finite temperature
has dramatic consequences on the macroscopic properties of \emph{f}-electron
systems including volume collapse \cite{Arvanitidis-Nature2003},
quenched magnetism \cite{ABDELMEGUID-PRL85}, onset of superconductivity
\cite{Matsumoto-Science2011,Okawa-PRL2010}, Kondo physics \cite{Dzero-PRL2006},
and quantum criticality \cite{Matsumoto-Science2011,Okawa-PRL2010}.
Despite mixed-valency being central to \emph{f}-electron physics,
our ability to directly probe this peculiar quantum electronic state
at high pressures is limited. 

EuO with its simple NaCl (B1) crystal structure is a model system 
to study valence effects upon lattice compression \cite{Wachter-1979}. Eu-containing compounds
are prototypical mixed-valent systems because Eu can display both
trivalent (as most Rare-Earths ions do) and divalent electronic states
at ambient conditions, the latter stabilized by a half-filled 4\emph{f}
orbital occupation ({[}Xe{]}4\emph{f}$^{7}$5\emph{d}$^{0}$\emph{6}s$^{2}$
configuration) \cite{Varma-RMP76}. Additionally, the ferromagnetic-semiconductor
character of EuO \cite{Wachter-1979} coupled with perfect spin-polarization
of electronic states near the Fermi level generated interest for possible
applications of EuO in Spintronics \cite{Schmehl-NatMat2007}. A dramatic,
three-fold increase in magnetic ordering temperature is observed under
applied pressures of up to $\approx$14 GPa (or at strained interfaces),
reaching a maximum $\mathrm{T_{C}\approx200 }$ K but decreasing at
higher pressures \cite{TISSEN-JETP87}. The relationship between crystal
structure, electronic structure and magnetic ordering temperature
has fueled much of the research in EuO over the last two decades,
with the question of Eu valency remaining key for a complete understanding
of this and other \emph{f}-electron mixed-valent systems.
In this \emph{Letter} we report direct measurements of electronic valence and crystal structure in Europium monoxide
(EuO) at pressures up to 90 GPa using x-ray absorption spectroscopy,
nuclear forward scattering and x-ray diffraction techniques. Below 40 GPa a complex, pressure-dependent valence is observed to fluctuate between Eu$^{2+}$ and  Eu$^{3+}$ states at a frequency  $f\gtrsim\Delta\mathrm{E}/\hbar\sim0.15$ PetaHertz where $\Delta\mathrm{E}$ is the  4\emph{f} bandwidth. At higher pressures we observe a novel reentrant valence transition in Eu ions into a lower valence state despite a concomitant 7\% volume collapse. 
Oftentimes, valence state is derived from structural data alone via bond valence sum rules \cite{Jayaraman-PRB74,BRESE-ActaCrystB-91,BROWN-ActaCrystB-85},
which relate interatomic distances and coordination number to valence. 
We demonstrate that bond-valence models fail to quantitatively describe the pressure-dependent mixed-valent states of EuO in a broad pressure range. The
results call for revisiting the vast literature on mixed-valent \emph{f}-electron
systems where electronic valence is inferred from structural data alone.

Polycrystalline samples of EuO were prepared as described in ref \cite{MAUGER-PR1986}.
X-ray diffraction (XRD), x-ray absorption near-edge spectroscopy (XANES) and
nuclear forward scattering (NFS) experiments under pressure were performed
at beamlines 16-BM-D, 4-ID-D and 3-ID-D of the Advanced Photon Source, Argonne National Laboratory, respectively. Sample loading into the diamond anvil
cell was carefully done in an Argon atmosphere to prevent oxidation. Pressure was calibrated in situ by the Ruby luminescence method  \cite{Syassen-Ruby-2008}. Additional details
of the experimental methods are presented below and in the supplemental material \cite{supplemental}. 

\begin{figure}
\includegraphics[scale=0.3]{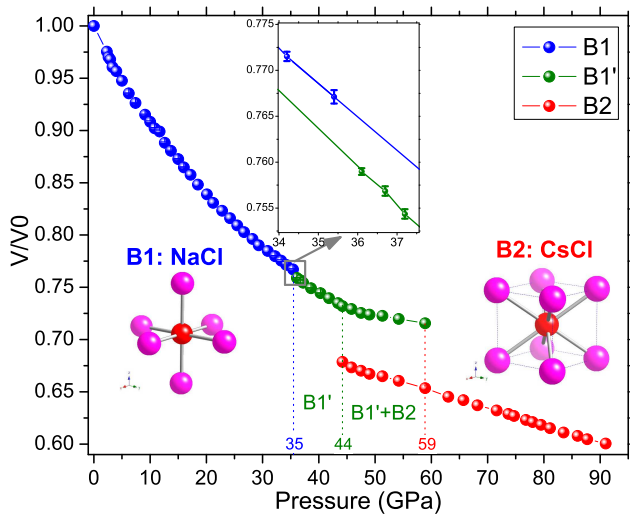}

\caption{Pressure-volume dependence of EuO up to 92 GPa obtained from Rietveld refinements
of XRD data. The inset panel shows
a modest iso-structural volume collapse at about 35 GPa (B1\textquoteright{}
phase). From 44 to 59 GPa a mixture of NaCl (B1) and CsCl (B2) phases is present.
Above 59 GPa a CsCl structural phase is homogeneous over the sample
volume. Two schematic figures show the local coordination in NaCl
and CsCl phases.}
\end{figure}

In view of the limited pressure range of previous studies, together
with conflicting reports on the pressure dependence of its lattice
parameter \cite{Jayaraman-PRB74,HEATHMAN-JAC1995}, we carried out
high-precision XRD measurements to determine the
pressure-volume relation in EuO up to 92 GPa (Fig. 1). We used
Neon as pressure-transmitting medium and diffraction peaks from reference Gold
particles for {\it in-situ} pressure calibration. A very modest volume collapse ($\approx$ 0.5\%) occurs at $\approx$ 35 GPa with no change in crystal structure (NaCl- B1), likely related to electronic instabilities as interpreted by Jayaraman {\it et al.} \cite{Jayaraman-PRB74}. Starting at about 44 GPa a first-order structural transition to a CsCl (B2) structure takes place with detectable
coexistence of both phases over the sample volume up to 59 GPa, in
fair agreement with a previous report \cite{HEATHMAN-JAC1995}. After
reaching about 92 GPa, the pressure was released and the structural changes were
observed to be reversible. XRD results were previously used to argue,
based on bond-valence sum rules \cite{BRESE-ActaCrystB-91,BROWN-ActaCrystB-85},
that the valence of Eu in  EuO dramatically increases as a function of pressure
with a first order valence transition towards a Eu$^{3+}$ state taking
place at the modest isostructural volume collapse resulting in a 2.5+
valence at about 35 GPa \cite{Jayaraman-PRB74,ZIMMER-PRB84}. While
a volume collapse could in principle be a signature of a sizable increase
in valence, just the opposite takes place at the structural phase
transition in EuO, as discussed below. 

\begin{figure}
\includegraphics[scale=0.3]{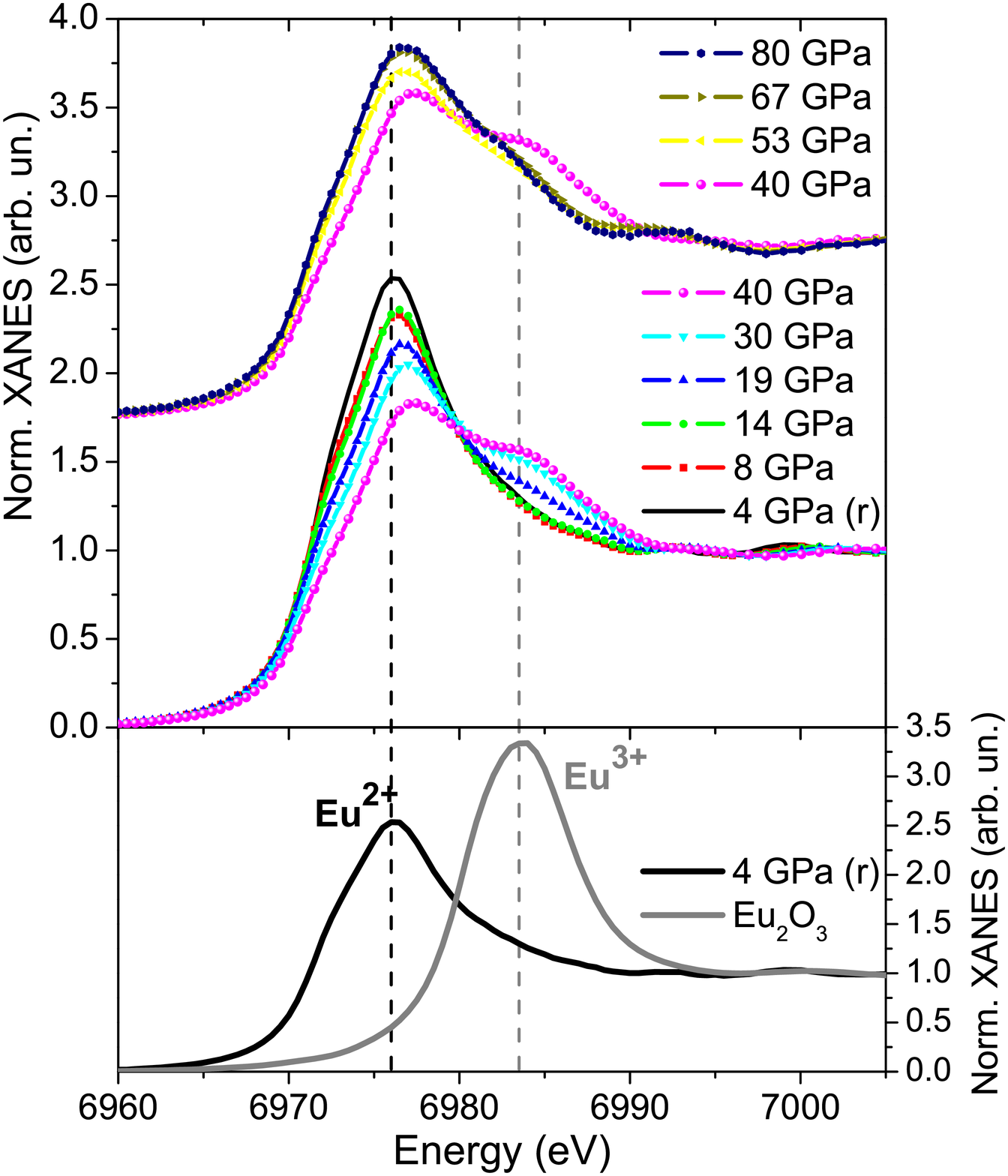}

\caption{Eu L$_{3}$ XANES  measurements on
EuO up to 80 GPa. The bottom panel compares spectra for Eu$^{2+}$ and Eu$^{3+}$ ions in EuO
and Eu$_{2}$O$_{3}$, respectively. An $\approx$8eV threshold energy
difference unequivocally distinguishes the two valence states. The
top panel shows XANES spectra in the 4 to 40 GPa pressure range where
fractional {[}P<14 GPa{]} and mixed valence {[}14<P<40 GPa{]} states
are observed, with translated spectra showing the reentrant valence
behavior at higher pressures {[}45-60 GPa{]} coincident with the volume
collapse transition. The 4 GPa spectrum was taken upon pressure release
indicating reversibility of the electronic transition.}
\end{figure}

\begin{figure}
\includegraphics[scale=0.29]{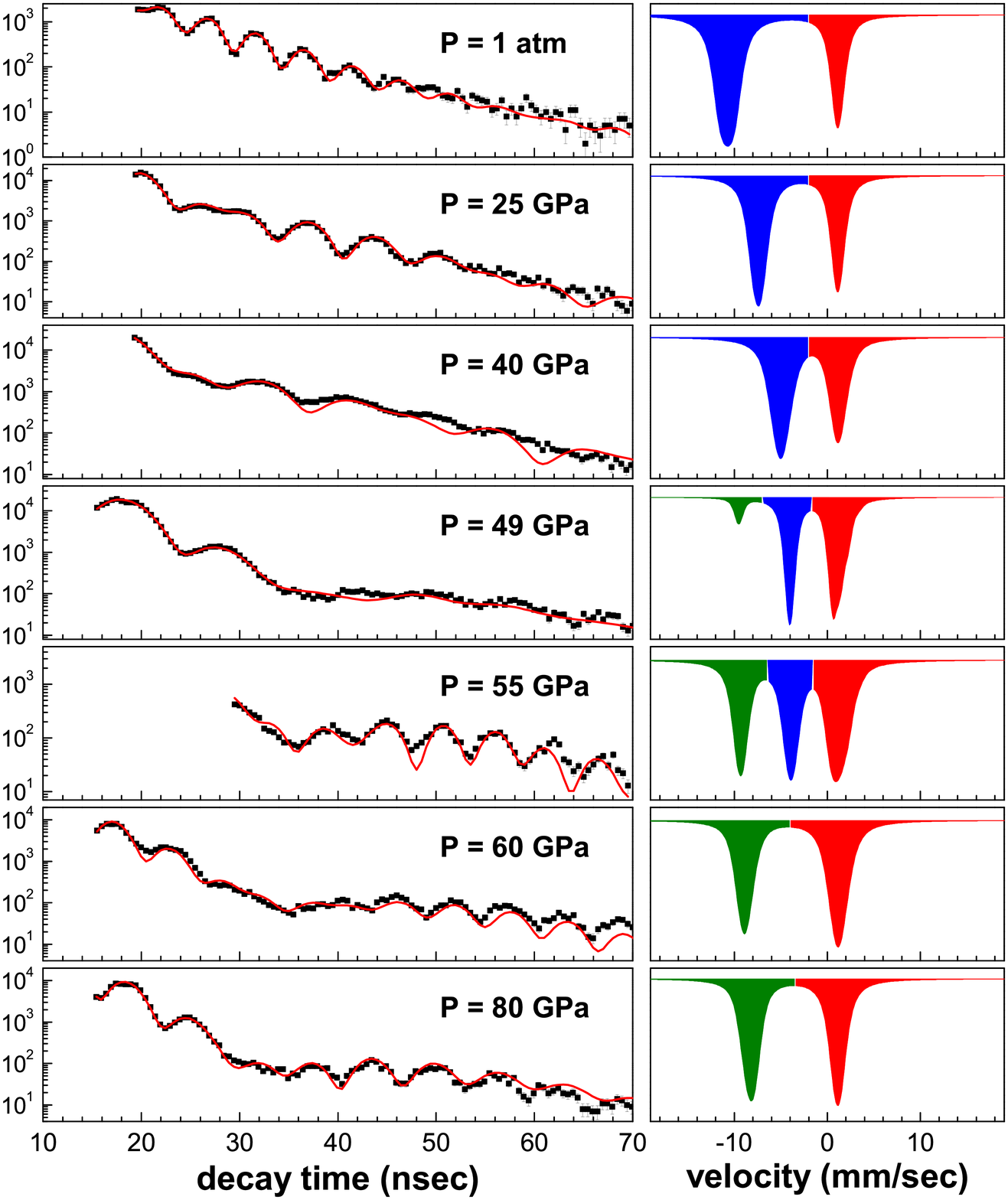}

\caption{Synchrotron Mössbauer spectra of EuO under pressure, collected using
an Eu$_{2}$O$_{3}$ sample at ambient pressure as reference, are
shown in the time domain in the left panel together with fits to the data. Fit results were used to simulate spectra in the
energy domain shown in the right panel. Red peaks correspond to the Eu$_{2}$O$_{3}$ reference located
after the diamond anvil cell, blue peaks correspond to the B1 phase
of EuO, and green peaks to the B2 phase. A complete dataset for
all pressures is given with the supplemental material \cite{supplemental}.}
\end{figure}

XANES data were collected in order to directly probe the electronic
structure of EuO as a function of lattice contraction. A previous
report \cite{Souza-Neto-PRL2009} provides evidence for a direct correlation
between Eu 4\emph{f}-5\emph{d} electronic mixing and the increase
in ferromagnetic ordering temperature that takes place in EuO up to
$\approx$14 GPa. Despite the presence of \emph{f}-\emph{d} mixing
no evidence of mixed valency was found in this pressure regime where
a stable, fractional (non-integer) occupation of 4\emph{f} orbitals
is found instead. The difference in excitation threshold for a 2\emph{p}$_ {3/2}$
\textrightarrow{} 5\emph{d} electronic transition (Eu L$_{3}$ absorption
edge) in 4\emph{f}$^{7}$5\emph{d}$^{0}$ (Eu$^{2+}$) and 4\emph{f}$^{6}$5\emph{d}$^{1}$
(Eu$^{3+}$) configurations is $\approx$8.0 eV, as shown in Fig.  2. While standards with known valence state, such as Eu$_{2}$O$_{3}$,
can be used to determine the degree of mixed valency an accurate estimate
must also consider the effects of crystal structure upon the XANES
spectra %
\footnote{Using the lattice parameters determined by X-ray diffraction, we used
first principles methods \cite{Souza-Neto-PRL2009,Joly-PRB01} to
simulate XANES spectra as a function of pressure for both Eu$^{2+}$
and Eu$^{3+}$ configurations. These reference theoretical spectra
were combined to match the experimental data and retrieve the ratio
of 2+/3+ valence states. This method takes into account the different shape of XANES spectra corresponding to the two valences, which is ignored in the method
of fitting the experimental data with Gaussian peaks \cite{Rohler-EuO84}.%
}. Between 14 and 40 GPa Eu is found to be in a mixed valence state,
reaching 2.21+ at 40 GPa. The fractional 4{\emph f} occupancy below $\approx$14
GPa, and the degree of mixed valency above this pressure agree well
with previous work \cite{ZIMMER-PRB84,EYERT-SSC86,Rohler-EuO84,TISSEN-JETP87}
where the valence of Eu was explored in the context of transport and
magnetic properties. The appearance of mixed valency appears to coincide
with the downturn in ferromagnetic ordering temperature\cite{TISSEN-JETP87}, 
a result of an increased fraction of  J=|L+S|$\approx$0 Eu$^{3+}$ ions.
Remarkably, at higher pressures a \emph{lower} valence state is abruptly
recovered concomitant with the B1\textrightarrow{}B2 structural phase
transition; i.e., the $\approx$7\% volume collapse is accompanied
by a \emph{decrease} in Eu valence. An increase in Eu-O bond length in the
high-pressure B2 phase as a result of the change in Eu coordination
number from N=6 (NaCl) to N=8 (CsCl) allows for the reentrant valence
transition to occur despite the sizable macroscopic volume contraction.

The pressure dependence of the Eu valence was verified with the NFS
technique\cite{ALP-NIM95}, which probes valence through the Mössbauer
isomer-shift (IS), namely, the change in \emph{s}-electron density at the nucleus
of a Mössbauer isotope as a result of changes in 4\emph{f} electron occupation
\cite{Varma-RMP76,Klein-SSC76}. An accurate description of valence
from NFS requires taking into account the compression of \emph{s}-like wave functions induced
by lattice contraction\cite{Klein-SSC76,ABDELMEGUID-PRB1990}, unrelated to changes in valence (see supplemental material \cite{supplemental} for details). The NFS data in Fig. 3 indicate that the
Eu$^{2+}$ valence changes towards Eu$^{3+}$ below 50 GPa, in fair
agreement with previously reported Mössbauer data \cite{ABDELMEGUID-PRB1990}
\footnote{Note that the correspondence between IS and valence in figure S5b of \cite{supplemental} is only directly applicable when both sample and reference are at ambient pressure. }.
At pressures above the structural phase transition, however, a near Eu$^{2+}$
state is recovered in good agreement with the XANES. Moreover, the
presence of a single resonance in the Mossbauer spectra unequivocally
demonstrates that the compound displays a spatially homogeneous Eu
valence (i.e. all Eu sites are equivalent) below and above the structural
phase transition. The exception is the region of phase coexistence
between 44 and 59 GPa where the appearance of a second resonance is
indicative of a spatially inhomogeneous Eu valence.

\begin{figure}
\includegraphics[scale=0.32]{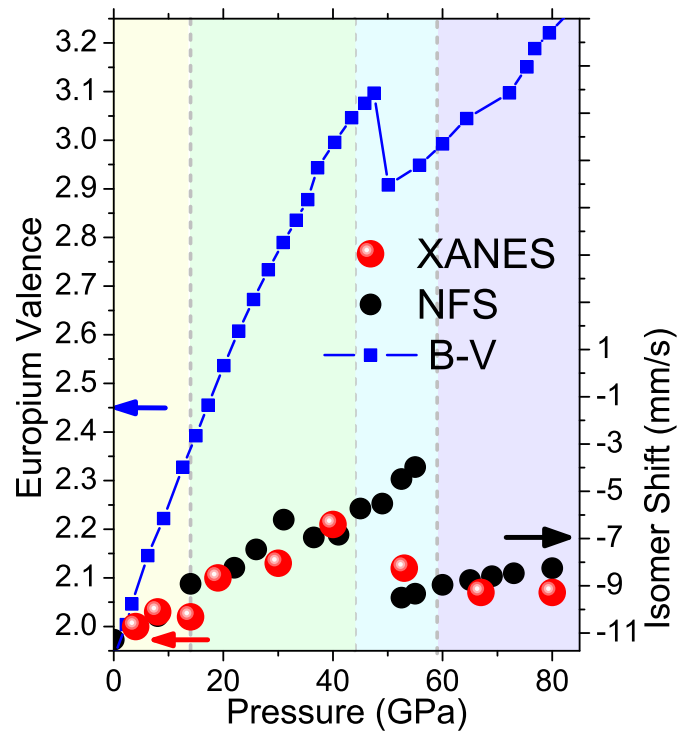}

\caption{ Europium valence determined by XANES, NFS and bond-valence-parameterization is shown in the left axis. The IS determined by NFS is shown in the
right axis. Valence from NFS is obtained after volume correction based on the B1 phase compressibility 
\cite{ABDELMEGUID-PRB1990}.
XANES and NFS results are in good agreement, but at odds
with the bond-valence-parameterization method. Colors differentiate the four different valence regimes  as described in the text.}
\end{figure}

While XANES spectra discretely present both Eu$^{2+}$ and Eu$^{3+}$
valence states at each pressure, the Mössbauer IS from NFS
shows a single, intermediate valence between Eu$^{2+}$ and Eu$^{3+}$
states. This is understood based on the different time scales of XANES
and NFS experiments. The probing time of about fifty attosec (core-hole lifetime of 5x10$^{-17}$
sec) in XANES spectroscopy at the Eu L$_{3}$ absorption edge (6.97
keV) is much faster than the 1.4x10$^{-8}$ sec lifetime of the $^{151}$Eu
nuclear excited state \cite{Haskel-PRB1998}. While a mixed-valence
state can be expressed as a superposition of 4\emph{f}$^{7}$ and
4\emph{f}$^{6}$ wavefunctions at a given instant, fluctuating valence
states are present due to the characteristic frequency defined by
the non-zero 4\emph{f} bandwidth \cite{Varma-RMP76}. At ambient pressure
a bandwidth of about 0.1 eV corresponds to a valence lifetime of $\approx$6x10$^{-15}$
sec. Consequently, the Mössbauer experiment sees a fast fluctuating
valence \cite{Varma-RMP76} as a static average of both states while
the XANES process is fast enough to separately probe both states.
Interestingly, the proximity in the lifetime of the fluctuating valence
state and lattice fluctuations ($\approx10^{-13}$) may influence
how the magnetism \cite{TISSEN-JETP87,Souza-Neto-PRL2009,EYERT-SSC86,ABDELMEGUID-PRB1990,Arnold-PRL2008}
and conductivity \cite{ZIMMER-PRB84,Arnold-PRL2008,DiMarzio-PRB87}
of EuO respond to pressure as the 4\emph{f} electronic bandwidth changes
with lattice contraction.

The general approximation \textquotedblleft{}bond length is a unique
function of bond valence\textquotedblright{} \cite{BRESE-ActaCrystB-91}
is widely used to determine a material\textquoteright{}s valence based
on crystal structure. The prediction of valence by these methods \cite{Varma-RMP76,Jayaraman-PRB74,BRESE-ActaCrystB-91,BROWN-ActaCrystB-85}
is done using a linear combination of lattice constants as a function
of pressure considering the ionic radius of, for example, Eu$^{2+}$
and Eu$^{3+}$ ions (which differ by about 10\%). We used the parameterization
method of \cite{BRESE-ActaCrystB-91} to calculate the valence based
on bond distances from XRD data (see supplemental material \cite{supplemental} for details).
The values of valences for EuO as a function of lattice parameter
are shown in Fig. 4, where we compare the valence determined by
this bond-valence parameterization \cite{BRESE-ActaCrystB-91} with
results from XANES and NFS experiments. It is clear that the over simplifications of the bond-valence method \cite{BRESE-ActaCrystB-91}, which determines valence based solely on atomic structure, give rise to
inaccurate results. For example, any nonlinear dependence of volume on valence,
together with changes in the degree of ionicity, would undermine the
accuracy of this method \cite{Varma-RMP76}.

In summary, we have shown that EuO presents four pressure regimes related to fractional and mixed-valent behaviors, which are highlighted here due to their important connection with
the magnetic and transport properties under pressure. Below 14 GPa
a well-defined quantum state of fractional 4\emph{f} (and 5\emph{d})
occupation evolves with pressure, resulting in a continuous increase in magnetic ordering temperature\cite{Souza-Neto-PRL2009}. Between 14 and 44 GPa the Europium atoms display a mixed-valent state composed of discrete Eu$^{2+}$ and Eu$^{3+}$ states, homogenously distributed over volume and fluctuating with a characteristic frequency determined by the 4\emph{f} bandwidth ($f\gtrsim0.15$ PHz). This leads to a continuous decrease in magnetic ordering temperature as a result of the J $\approx0$  state of Eu$^{3+}$ ions. From 44 GPa to 59 GPa, coexistence of NaCl and CsCl structures results in a spatially inhomogeneous valence state, the CsCl structure displaying a reentrant, nearly Eu$^{2+}$, valence state which becomes spatially homogeneous above 59 GPa. It remains to be seen if the reentrant valence transition is associated with a strengthening of magnetic interactions in EuO
at high pressures, interactions that were otherwise weakened by the
mixed-valent state \cite{EYERT-SSC86,TISSEN-JETP87}.
Most importantly we show that widely used bond-valence methods fail
to quantitatively describe essential features of the complex electronic valence behavior
of EuO under pressure. This highlights the need for probing both structure
and electronic states directly when valence information is sought
in mixed-valent systems.

\begin{acknowledgments}
We are grateful to James S. Schilling and Yves Petroff for commenting
on the manuscript. Work at Argonne is supported by the U.S. Department
of Energy, Office of Science, Office of Basic Energy Sciences, under
Contract No. DE-AC-02- 06CH11357. HPCAT is supported by CIW, CDAC,
UNLV and LLNL through funding from DOE-NNSA, DOE-BES and NSF.
\end{acknowledgments}
\bibliographystyle{apsrev}

\end{document}